\documentstyle[pra,aps,twocolumn,epsfig]{revtex}

\begin{document}

\title{Laser Induced  Condensation  of Trapped Bosonic Gases}

\author{L. Santos$^{1,2}$, M. Lewenstein$^{2}$, J. I. Cirac$^3$,
and Y. Castin$^4$}

\address{(1) Departamento de F\'\i sica Aplicada, Universidad de
Salamanca, 37008 Salamanca, Spain \\
(2) Institut f\"ur Theoretische Physik, Universit\"at Hannover,
 D-30167 Hannover,
Germany\\
(3) Institut f\"ur Theoretische Physik, Universit\"at Innsbruck,
 A--6020 Innsbruck, Austria\\
(4) Laboratoire Kastler Brossel, Ecole Normal Sup\'erieure,
 F-75231 Paris Cedex 05, France}

\maketitle

\begin{abstract}
We demonstrate that an appropriate sequence of laser pulses allows to
condense a gas of trapped bosonic atoms into an arbitrary trap level.
 Such condensation is robust, can be achieved in
experimentally feasible traps, and may lead to multistability and
hysteresis phenomena.
\end{abstract}



Laser cooling has led  to spectacular results in
recent years \cite{nobel}. So far, however, it has not allowed to reach
temperatures for which quantum statistics become important.
In particular, evaporative  cooling is used to obtain
Bose-Einstein condensation of trapped gases \cite{BEC}.
Nevertheless, several groups are pursuing the challenging goal of
condensation via all--optical means \cite{Salomon,Ertmer,Mlynek}.

In traps of size larger than the inverse
wavevector, $k_L^{-1}$, the temperatures required for condensation are
below, or of the order of, the photon recoil energy,
$E_R=\hbar\omega_R=\hbar^2 k_L^2/2m$, where $m$ is the atomic mass.
There exist several laser cooling schemes to reach such temperatures
\cite{VSCPT,Raman}. They exploit  single atom
``dark states'', i.e. states which cannot be excited by the
laser, but can be populated via spontaneous emission. The main
difficulty in applying dark state cooling  for dense
gases is caused by light reabsorbtion. Unfortunately, these
states are not dark with respect to the photons spontaneously emitted by
other atoms. Thus, at sufficiently high densities, dark state cooling
may cease to work, since multiple reabsorptions can increase the system
energy by several recoil energies per atom
\cite{Sesko,Dalibard,Burnett,Ellinger}. In particular, laser induced
condensation is feasible only if the reabsorbtion probability is smaller
than the inverse of the number of energy levels accessible via
spontaneous emission processes \cite{Dalibard}.

Several remedies to the reabsorption problem have been
proposed. First, the role of reabsorptions increases with the
dimensionality. If the reabsorption cross section for trapped
atoms is the same as in free space, i.e. $\simeq 1/k_L^2$, the
reabsorptions should not cause any problem in 1D, have to be carefully
considered in 2D, and forbid condensation in 3D. Working with quasi-1D
or -2D systems is thus a possible way to reduce the role of
reabsorptions \cite{Janicke}.  The most promising remedy
against  this problem employs the
dependence of the reabsorption probability for trapped atoms on the
fluorescence rate $\gamma$, which can be easily adjusted in
dark state cooling \cite{reab}. In particular, in the so called {\em
Festina Lente} limit, when $\gamma$ is much smaller than the trap
frequency $\omega$ \cite{Festina}, the reabsorption processes in which
the atoms change energy and undergo heating are suppressed.
However, neither collective cooling schemes in traps of realistic size
have been investigated in this limit, nor has it been shown that
laser induced condensation is possible.

In this Letter we present such investigation. First, we formulate the
Master Equation (ME) that describes Raman  cooling in the Festina Lente
limit using coarse
graining in time. Cooling  is dynamical, and consists of
sequences of pairs of laser pulses inducing stimulated and
spontaneous Raman transitions between the two electronic levels
of trapped atoms, $|g\rangle$ and $|e\rangle$. The stimulated absorption
pulses induce  the energy selective transition
$|g\rangle\to|e\rangle$ that depopulates all motional states except the
dark ones; the spontaneous emission pulses are non-selective, and repump
the atoms from $|e\rangle$ to $|g\rangle$ populating all accessible
motional states. We simulate the dynamics generated by the ME in 3D, and
show that: i) laser induced condensation into an
arbitrary trap level is possible;  an arbitrary trap level
may be made dark; ii) condensation is robust with respect to changes of
physical parameters; dark states do not have to be completely dark;
iii) multistability and hysteresis occur when the parameters
undergo large changes. In the limit of large number of atoms analytic
solutions of the ME are found.

We consider $N$ bosonic atoms with two levels
$|g\rangle$ and $|e\rangle$ in a non-isotropic trap with the frequencies
$\omega^g_{x,y,z},\omega^e_{x,y,z}$ different for the ground and the
excited states, and non-commensurable one with another. This assumption
simplifies enormously the dynamics of the spontaneous emission processes
in the Festina Lente limit. We use the coarse graining in time and
describe  variations  of the atomic state after one
absorption and one spontaneous emission pulse. After such cooling cycle
all atoms are in the ground internal state described by the density
matrix $\rho(t)$. This matrix is diagonal in the Fock representation
corresponding to the bare trap levels.  In order to derive
the ME, we separate the effects of laser cooling
from the ones due to  atom--atom collisions. The latter can
be described by a quantum kinetic ME, which has been
studied in Ref. \cite{gard}. In this paper we 
concentrate on the laser cooling process. Thus, our results are
valid in the case when  collision processes are slow compared to laser cooling.

For the stimulated Raman transitions, we assume
weak pulses of duration $\tau_{abs}$. Their effects can thus be
described by second order perturbation theory (formally that corresponds
to one atom excited at most),
\begin{eqnarray}
&&\rho(t+\tau_{abs})=\rho(t)-\sum_{lm}\Gamma^{abs}_{lm}g_m^{\dag}g_m\rho(t)
\nonumber \\
&&-\sum_{lm}\Gamma^{abs}_{lm}\rho(t)g_m^{\dag}g_m+
2\sum_{lm}\Gamma^{abs}_{lm}e_l^{\dag}g_m\rho(t)e_l g_m^{\dag},
\label{abs}
\end{eqnarray}
where $g_m, g_m^{\dag}$ ($e_l, e_l^{\dag}$) are bosonic annihilation and
creation operators of atoms in the ground (excited) internal state and
in the trap level $m=(m_x,m_y,m_z)$ [$l=(l_x,l_y,l_z)$]. The
absorption probabilities $\Gamma^{abs}_{lm}$ describe transitions from
the ground state level $m$ to an excited state level $l$, and depend on
Raman laser pulses. For instance, if the
Raman transition is characterized by the maximal (effective) Rabi
frequency $\Omega_0$, temporal envelope $f(t)$, wavevector ${\bf k}_L$,
frequency $\omega_L$, and detuning $\delta=\omega_L-\omega_a$, where
$\omega_a$ is the internal levels energy difference, the probabilities
are given by
\begin{equation}
     \Gamma_{lm}^{abs}=\frac{\Omega_0^2}{8}
|\langle l|e^{i{\bf k}_L{\bf x}}|m\rangle|^2
|\tilde
 f(
\delta-\omega_l^e+\omega^g_m)|^2,
\end{equation}
where the first term is  Franck-Condon factor, and $$\tilde
f(\delta)=\int_{-\infty}^{+\infty}f(t)e^{-i\delta t}dt,$$ is a function
peaked at $\delta=0$ with width of the order of $1/\tau_{abs}$. In the
following we use Gaussian pulses with
$f(t)=\exp(-t^2/\tau_{abs}^2)$ and  $\omega\tau_{abs}>1$, so that only 
resonant  $\Gamma_{lm}^{abs}$ are relevant 
($|\delta-\omega_l^e+\omega^g_m|<\tau_{abs}^{-1}$). 
On the other hand, $\Omega_0\tau_{abs}<1$, so that only a small fraction 
of atoms is excited by the absorption pulse, which ensures the validity of
the perturbative approach. In Eq. (\ref{abs}) two kinds of terms
are omitted, since their contributions
vanish after the repumping process: i) the
first order terms, which correspond to coherences with
respect to internal levels (off-diagonal matrix elements between the
states with one excited, and no excited atoms); these
terms  are destroyed after spontaneous emission; ii)
coherences between different excited trap levels; those terms are
neglected since spontaneous emission in the Festina Lente
limit is purely diagonal in the $l$ index; thus $l\ne l'$ coherences
are destroyed in the quantum jump down to the ground state, and do not
influence the dynamics of diagonal matrix elements \cite{cohe}.

The repumping pulses have constant amplitude and duration $\tau_{sp}$
long enough to depopulate totally the excited states. In accordance with
the Festina Lente limit, we will assume that (collective) spontaneous
emission rates are small in comparison to the trap frequency
\cite{Festina}. This allows to perform a secular approximation in the
master equation describing this process, which leads to the following
equation after spontaneous emission has taken place ($\tau_{ch}=\tau_{abs}
+\tau_{sp}$):
\begin{eqnarray}
&&\rho(t+\tau_{ch})=\rho(t)-
\sum_{lm}\Gamma^{abs}_{lm} \left [
g_m^{\dag}g_m\rho(t)+\rho(t)g_m^{\dag}g_m \right ] \nonumber \\
&&+4\int_0^{\infty}d\tau \sum_{mnl}\left[\Gamma^{sp}_{ln}\Gamma^{abs}_{lm}
g^{\dag}_n
e^{-\sum_{k}\Gamma^{sp}_{kl}g_kg_k^{\dag}\tau}g_m\right.\nonumber \\
&&\left.\rho(t)g_m^{\dag}
e^{-\sum_{k}\Gamma^{sp}_{kl}g_kg_k^{\dag}\tau}g_n\right],
\label{sp}
\end{eqnarray}
The last term in Eq. (\ref{sp}) describes the integral over all
possible times $\tau$ in which the quantum jump from the excited state
$l$ to the ground state $n$ occurs. The amplitude of the excited state
is damped during the time $\tau$ with the collective rate
$\sum_k\Gamma^{sp}_{kl} g_kg_k^{\dag}$, which evidently contain Bose
enhancement factors, i.e. is proportional to occupation numbers plus one
of the corresponding ground trap levels. The integral over $\tau$ can be
extended to $\infty$ since $\tau_{sp}$ is long enough. The spontaneous
emission rates are
\begin{equation}
\Gamma^{sp}_{nl}=\gamma \int d\Omega W(\Omega)
|\langle n|e^{ik_a{\bf n}(\Omega){\bf x}}|l\rangle|^2,
\label{raTE}
\end{equation}
where $2\gamma$ is the single atom (effective) spontaneous emission rate
(i.e. spontaneous Raman transition rate, controllable in experiments),
integration extends over the solid angle $\Omega$,
$W(\Omega)$ describes dipole radiation pattern, ${\bf n}(\Omega)$ is a unit
vector in the $\Omega$ direction, and $k_a=\omega_a/c$.

Eq. (\ref{sp}) describes an elementary cooling step, i.e.
maps a diagonal density operator $\rho(t)$ onto a diagonal
$\rho(t+\tau_{ch})$, with all atoms
in the ground state. It can thus be simulated using standard
Monte Carlo procedures. We have performed such simulations for $N=1$ up
to $N=500$ atoms, in various dimensions, and for various cooling
strategies. Franck-Condon factors and trap frequencies  can
be efficiently approximated using the states of an isotropic trap of
frequency $\omega$. We concentrate on 3D cooling, beyond the Lamb-Dicke
limit, i.e. for the traps for which the Lamb-Dicke parameter
$\eta=\sqrt{E_R/\hbar\omega}$ is larger than one.  Due to
memory storage and calculation times, our numerical simulation has to be
restricted to values of $\eta\alt 2$. In that case the atom
(having initially an energy of the order of few $E_R$) may increase
its trap energy level in the spontaneous emission process by energies
$\sim E_R$, and non-standard cooling schemes have to be used to avoid
such heating effects \cite{Morigi}. Generalization of the approach of
Ref. \cite{Morigi} allows to cool dynamically (i.e. by changing
absorption laser pulses appropriately) {\it individual atoms} to arbitrary
trap levels \cite{Santos}.

The full dynamical cooling cycle 
must contain sequences of absorption pulses of
appropriately chosen frequencies. The Fourier bandwidth of the pulses
can be smaller than $\omega$, so that various resonance conditions may
be employed. We use the following types of pulses: i) {\it confinement
pulses}: spontaneous emission may increase each of the quantum numbers
$m_{x,y,z}$ by $O(\eta^2)$. In $D$-dimensions pulses with detuning
$\delta = -D\hat \eta^{2} \omega$, where $\hat\eta^2$ is the closest
integer to $\eta^2$, have thus an overall cooling effect, and confine
the atoms in the energy band of $D$ recoils. The use of two sligthly
detuned confinement pulses is recommended;
ii) {\it dark-state cooling pulses}: these pulses should fulfill dark
state condition for a selected state to which the cooling should occur;
iii) {\it sideband and auxiliary cooling pulses}: in general, dark state
cooling pulses might lead to unexpected trapping in other levels. In
order to avoid it, auxiliary pulses that empty undesired dark states and
do not empty the desired dark state are needed. For cooling into the
ground state, for instance, the sideband cooling pulse with $\delta =
-\omega $ is used; iv) {\it pseudo-confining pulses}: with the  use
of pulses i)--iii) cooling is typically very slow. In order
to shorten cooling time we use pulses with  $\delta = -3
\eta^{2}\omega/2$ and $\delta = -\eta ^{2}\omega$, which 
pseudo-confine the atoms below
$n=3\eta ^{2}/2$ and $n= \eta ^{2}$.

\begin{figure}[ht]
\begin{center}\
\epsfxsize=7.0cm
\hspace{0mm}
\psfig{file=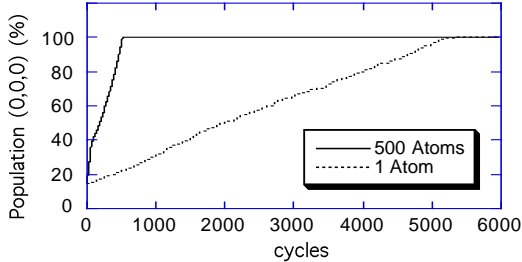,width=7.0cm}\\[0.1cm]
\caption{Population of  $(0,0,0)$ as a function of number of cycles, each
consisting of a
sequence of absorption pulses 
with  $s$=-12,-6,-4,0,-13,-7,-5,-1, and repumping pulses. For all the
absorption pulses $A_{x,y,z}=1$,
 except for $s=0$ for which $A_{z}=-2$. The initial
distribution is thermal with mean $\langle n\rangle =6$, and 
 $\eta=2.0$. Dotted (solid)
lines represent the case of $N=1$ ($500$) atoms.}
\label{fig:1}
\end{center}
\end{figure}
Let us now consider possible dark state conditions, determined by the
total probabilities of emptying a given level $m$,
$\Gamma_m=\sum_l\Gamma_{lm}^{abs}$. We consider three Raman
transitions induced by laser pairs propagating in directions $x$, $y$,
and $z$ characterised by three different effective Rabi frequencies
$\Omega_{0}f(t)A_{j}$, where $A_{j=x,y,z}$ account for difference of
intensities or dephasing between the lasers. As we discussed in detail
in Ref. \cite{Santos} the dark states may appear due to two reasons:
vanishing of all Franck-Condon coefficients for three lasers, or
destructive interference effects between the three lasers. The first
type of conditions can be achieved by choosing appropriate detuning
$\delta=s\omega$ with integer $s$: for instance, ground state
$m=(0,0,0)$ is dark with respect to side-band cooling pulses with
$s=-1$. Since the Franck-Condon coefficient $\langle
m_x+s|e^{ik_Lx}|m_x\rangle$ vanishes for $m_x=1$ provided $\eta^2=s+1$,
and analogous property holds for $y$ and $z$, the state $(1,1,1)$ can be
made dark provided $\eta^2$ is integer, and $s=\eta^2-1$. Similar,
conditions can be found for $m=(2,2,2)$. The second type of dark state
conditions correspond to use of resonant absorption with $s=0$. Chosing
for instance $A_{x}=1$ and $A_{y,z}$ such that $1+A_y+A_z=0$ makes all
states $(m,m,m)$ dark.

In Fig. 1 we present our result for ground state cooling of 1 and 500
sodium atoms in 3D trap with $\eta=2$, using 20 3D-energy
shells (i.e. 1771 trap levels). The initial state of the
system corresponds to mean energy $6\hbar \omega$, and is the same for
all figures. The pulse sequence is $s=$ -12,-6,-4, 0, -13, -7, -5,-1.
$A_{x,y,z}=1$, except for $s=0$, for which $A_{x,y}=1, A_z=-2$. Pulses 1
and 5 are confining, 2,3,6 and 7 pseudo-confining, and 4 and 8 are dark
state cooling pulses. The many body effects introduce one very important
element to the dynamics: Bose enhancement factors, that speed up the
dynamics enormously. The time scale is such that each cooling cycle must
be, say few times  longer than $2\pi/\omega \simeq 10^{-4}$s \cite{timesc}.
Then the function $\tilde f$ is  sufficiently narrow to neglect
non-resonant transitions. Therefore 1000 cycles correspond
to about 1s. Keeping $\Gamma$ fixed,
cooling of one atom requires here few seconds,
whereas collective cooling takes about $0.1$s. After achieving
condensation with 500 atoms, confining pulses can be avoided, a single
dark state pulse can keep the atoms in the condensed state.

\begin{figure}[ht]
\begin{center}\
\epsfxsize=7.0cm
\hspace{0mm}
\psfig{file=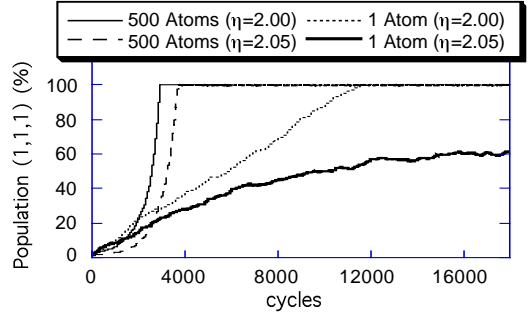,width=7.0cm}\\[0.1cm]
\caption{Population of $(1,1,1)$ as a function of number of cycles, 
each consisting of a
sequence of absorbtion pulses with $s=$-12,-6,-3,3,-13,-7,-4,-2, and 
$A_{x,y,z}=1$. The initial distribution is thermal with mean
$\langle n\rangle =6$. The cases of  $\eta=2.0$ and $2.05$, $N=1$ and $500$
are depicted.}
\label{fig:2}
\end{center}
\end{figure}
In Fig. 2 we show results for cooling into the state $(1,1,1)$ using the
sequence $s=$-12,-6,-3,3,-13,7,-4,-2, with $A_{x,y,z}=1$. Here the pulse
4 is the dark state pulse ($\eta^2=s+1$), the other are either confining
or auxiliary. First, note that when dark state condition is fulfilled
exactly ($\eta=2$), cooling of a single atom to $(1,1,1)$, although
slow, is possible. This cooling mechanism is, however, very fragile. A
tiny perturbation of the dark state ($\eta=2.05$) makes efficient
cooling impossible. This conclusion does not hold for many atoms though.
Quantum statistics helps to achieve 100\% condensation that is robust
with respect to parameter changes; the results for $\eta=2$ or 2.05 are
undistinguishable, cooling is much shorter than in the 1 atom case, and
takes about 1s \cite{fini}.

The existence of various stationary states in our system suggest the
possibility of multistability and hysteresis effects
\cite{cirac}. Indeed, in Fig. 3 we investigate the same
cooling sequence as in Fig. 1, except that $s_{8}$=-2, and for pulse 4
$A_z$ varies adiabatically from $-1.94$ to $-0.08$ and back during 37200
cycles. For $A_z$ close to -2, the ground state is in this case dark for
$s=0$. As $A_z$ grows, at some point ($A_{z}=-2/3$) the
states $(1,0,1)$ and $(0,1,1)$ become dark. This occurs when the
destructive interference
\begin{equation}
\langle 0_x|e^{ik_Lx}|0_x\rangle
+\langle 1_y|e^{ik_Ly}|1_y\rangle
=- A_z\langle 1_z|e^{ik_Lz}|1_z\rangle
\label{inte}
\end{equation}
takes place. The system shows multistability and hysteresis:
the  transfer from $(0,0,0)$ to $(1,0,1)$ and $(0,1,1)$ occurs
for higher values of $A_z$ than vice versa.

\begin{figure}[ht]
\begin{center}\
\epsfxsize=7.0cm
\hspace{0mm}
\psfig{file=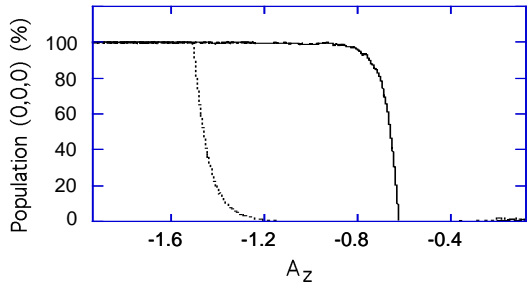,width=7.0cm}\\[0.1cm]
\caption{Hysteresis cycle in the condensation dynamics into $(0,0,0)$, for
a sequence of
pulses $s=$-12,-6,-4,0,-13,-7,-5,-2 with $A_{x,y,z}=1$
for all the pulses
except for the pulse $s=0$.   For $s=0$   $A_{z}=-1.94$ for the first 1200
pulses, $A_{z}$
increases
adiabatically during $18600$ cycles up to $-0.08$ (solid line).
In the next
$18600$ cycles
$A_{z}$ goes back to its original value $-1.94$ (dotted line).}
\label{fig:3}
\end{center}
\end{figure}


If most atoms are  condensed, confining
 pulses are no more needed. One can study then the stationary
limit of the ME with cooling pulses with fixed $s$. Amazingly, using the
Glauber's ${\cal}P$ representation and systematic $1/N$ expansion the ME
can be solved analytically. The results can be summarized as follows: i)
conditions for absorption and spontaneous emission rates that lead to
cooling into an arbitrary state are analytically obtained;
the system condenses into $n_0$--state iff 
$\tilde\Gamma_n\equiv \sum_m(\Gamma_{mn}^{abs}- \Gamma_{mn_0}^{abs}
\Gamma_{nm}^{sp}/\Gamma_{n_0m}^{sp})>0$ for all $n\ne n_0$.
 ii) the
dynamics exhibits two time scales: on a fast scale (several cooling
cycles), noncondensed modes behave as independent quantum harmonic
oscillators that approach thermal equilibrium (quantum
Ornstein-Uhlenbeck processes). The condensate mode is correlated to that
dynamics through the atom number conservation. The cooling time 
is of the order of $\max_n (\tau_{ch}/\tilde\Gamma_n)$. On a slow time scale
(that is $N$ times slower than the fast one) the dynamics is dominated
by self-transitions from the condensate and back. That
produces slow phase diffusion and slow decay of the two-time
correlation function of the condensed mode  with the rate 
$\simeq \sum_m \Gamma_{mn_0}^{abs}/2N\tau_{ch}$. The scattered
photon statistics is Poissonian.

Sumarizing, using the quantum ME in the Festina Lente limit, we have
demonstrated that  properly designed sequences of laser pulses
allow to condense a gas of trapped bosonic atoms into an arbitrary
state. Such condensation is robust, can be achieved in experimentally
feasible traps, and leads to multistability and hysteresis phenomena. We
have neclected in our approach atom--atom collisions.
Collisionally
induced population redistribution should not affect condensate in the
(collisionally modified) ground state $(0,0,0)$, since this state
corresponds to thermal equilibrium at very low temperatures.
Thermalisation mechanism might destroy condensates in excited states,
but that depends on time scales. Condensation requires 
seconds, but we have not attempted to optimize this
time. Optimal  times become shorter than collisional thermalisation time
if $N$ is not too large, and $\eta$
not too small.

We acknowledge  support from Spanish Direcci\'on General
de Investigaci\'on
Cient\'\i fica y T\'ecnica (Grant PB95-0955),  Junta de
Castilla y Le\'on
(Grant SA 16/98), and  Deutsche Forschungsgemeinschaft (SFB
407).


\begin{references}
\bibitem{cohe} We neglect also 
coherences with $\omega_m^g=
\omega_{m'}^g$, since due to selection rules the laser causes only one 
$m\rightarrow l$ transition.
\bibitem{timesc} $\tau_{abs}$ has to be such
that on average one absorption per elementary cooling step takes place.
That means that $\tau_{abs}$ has to be increased as the 
cooling approaches the final stages.
At the same time 
the pulses in the cooling cycle can be
limited to dark state pulses only. In effect the 
 duration of the full cycle remains more, or less 
constant, and represents well the time scale of cooling.
\bibitem{fini} The bosonic enhancement allows 
for condensation in unfavorable cases for $N\geq N_{crit}$, where
the critical number of atoms is quite sharply defined.
\end{references}
\end{document}